\documentclass[%
 reprint,
superscriptaddress,
unsortedaddress,
 amsmath,amssymb,
 aps,
]{revtex4-2}

\usepackage{float}
\usepackage[dvipsnames]{xcolor}
\usepackage{graphicx}
\usepackage{dcolumn}
\usepackage{bm}
\usepackage[colorlinks=true,linkcolor=blue,citecolor = blue,urlcolor=blue]{hyperref}

\newcommand{\hly}[1]{#1}

\begin{document}

\preprint{APS/123-QED}

\title{Magnetostatic response and field-controlled haloing in binary superparamagnetic mixtures}

\author{Andrey A. Kuznetsov}
\email{andrey.kuznetsov@univie.ac.at}
\affiliation{Computational and Soft Matter Physics, Faculty of Physics, University of Vienna, Kolingasse 14-16, 1090 Vienna, Austria}
\author{Sofia S. Kantorovich}
\affiliation{Computational and Soft Matter Physics, Faculty of Physics, University of Vienna, Kolingasse 14-16, 1090 Vienna, Austria}
\affiliation{Research Platform MMM Mathematics-Magnetism-Material, University of Vienna, Oskar-Morgenstern-Platz 1, 1090 Vienna, Austria}

\begin{abstract}
Nowadays, magnetoresponsive soft materials, based not simply on magnetic nanoparticles, but rather on multiple  components with distinct sizes and magnetic properties, both in liquid and polymeric carriers, are becoming more and more wide-spread due to their unique and versatile macroscopic response to an applied magnetic field. The variability of the latter is related to a complex interplay of the magnetic interactions in a highly non-uniform internal fields caused by spatial inhomogeneity in multicomponent systems. In this work, we present a combine analytical and simulation study of binary superparamagnetic systems, containing nanoclusters and dispersed single-domain nanoparticles, both in liquid and solid carrier matrices. 
We investigate the equilibrium magnetic response of these systems in wide ranges of concentrations and interaction energies. 
It turns out that, while the magnetisation of a binary solid can be both above and below that of an ideal superparamagnetic gas, depending on the concentration of the dispersed phase and the interparticle interactions, the system in a liquid carrier is highly magnetically responsive. In liquid, a spatial redistribution of the initially homogeneously dispersed phase in the vicinity of the nanocluster is observed -- the effect that is  reminiscent of the so-called ``haloing'' effect previously observed experimentally on micro- and milli-scales.
\end{abstract}

\maketitle


\section{Introduction}

Magnetic soft matter is a family of artificially synthesized materials 
based on a distributed system of magnetic particles embedded in a non-magnetic carrier matrix.
Notable members of the said family are ferrofluids~\cite{shliomis1974magnetic}, magnetorheological fluids~\cite{bossis2002magnetorheological}, ferrogels~\cite{weeber2018polymer} and magnetoactive elasomers~\cite{bellan2002field}.
The behavior and properties of these systems can be controlled using applied magnetic fields, 
which makes them highly attractive in various branches of nanotechnology and nanomedicine. 
Examples of applications include soft crawling robots~\cite{zimmermann2006modelling}, 
tissue engineering scaffolds~\cite{bock2010novel}, 
adaptive dampers and seals~\cite{abramchuk2007novel}, ferrofluid cooling systems~\cite{cherief2017parameters}, 
magnetic lubricants~\cite{huang2011study}, 
targeted drug delivery systems~\cite{tietze2013efficient}, 
magnetic hyperthermia of cancer~\cite{perigo2015fundamentals}  
and magnetic particle imaging~\cite{tay2018magnetic}. 

Modern methods of magnetic soft matter synthesis have achieved a tremendous success. 
In particular, particles can vary greatly in size and can have very different internal magnetic structure.
The common types of particles are single-domain ferro- or ferrimagnetic nanoparticles with linear sizes $\sim 10$ nm~\cite{rosensweig1985ferrohydrodynamics}, 
dense cluster of single-domain nanocrystals (magnetic ``nanoflowers''~\cite{bender2019supraferromagnetic} and ``multicore nanoparticles''~\cite{ludwig2014magnetic,socoliuc2022ferrofluids} with the size of the order of $\sim 10^2$ nm) 
and multi-domain microparticles with low or high coercivity~\cite{borin2019hybrid}.
Recently, multicomponent systems, which simultaneously employ several types of magnetic particles, attracted a lot of scientific attention.
For instance, these are hybrid elastomers containing both magnetically soft and hard microparticles~\cite{sanchez2019modeling}.
Reportedly, they allow for a much larger degree of a magneto-mechanical fine-tuning than analogous one-component systems~\cite{becker2018dynamic}.  
\hly{In Ref.~\cite{fischer2021}, an elastic sphere filled with magnetically saturated colloidal particles 
of two different sizes was considered  -- it was shown that for certain spatial arrangements of particles,
variation in the quantitative ratio between small and large particles  
can lead to qualitative changes in the system overall deformation response.} 
Another example are bimodal magnetorheological fluids,
which consist of magnetic microparticles submerged in a nanodispersed ferrofluid~\cite{magnet2012haloing}. 
They are considered to be an improved substitution for conventional magnetorheological fluids 
due to their superior colloidal stability and sedimentation behavior~\cite{viota2007study,lopez2010repulsive}. 
\hly{Recently, a novel type of binary ferrofluids containing a mixture of 
magnetically hard and magnetically soft nanoclusters was experimentally investigated in Ref.~\cite{khelfallah2023}.}
Even some samples of traditional ferrofluids are known to contain a fraction of large nanoclusters,
which results in a substantial alteration of their magnetic, mass-transport and rheological properties~\cite{buzmakov1996structure,rosensweig2019magnetorheological,pshenichnikov2012magnetophoresis,borin2011ferrofluid}.

The more the magnetic soft matter systems that contain two types of magnetic
components are developed, the clearer becomes the demand to understand the fundamental interplay between interactions of those components and their impact on the system overall magnetic response. 
The latter is of particular importance as it forms the basis for the efficient usage of these materials.  

Here, we will focus mainly on composite materials that are based on single-domain fine particles.
It is known that if the internal anisotropy energy of such particles is comparable or smaller 
than the energy of thermal fluctuations (which is common for iron oxide nanoparticles~\cite{rosensweig1985ferrohydrodynamics}),
then the average ensemble magnetization in zero field is zero. 
As the field increases, 
the magnetization will non-linearly and reversibly grow towards the saturation value.
Such behaviour is known as ``\textit{superparamagnetism}'' 
and corresponding materials can be referred to as ``\textit{superparamagnetic}''~\cite{bean1959superparamagnetism}.
While interactions in one-component superparamagnetic systems are very important, both in liquid~\cite{huke2004magnetic,ivanov2016revealing} and solid~\cite{elfimova2019static,radushnov2022structure} carriers, 
the situation becomes even more complex, if the material is multi-component. 
A clear evidence of this  are direct and inverse ferrofluid emulsions 
that are binary systems with only one magnetic component~\cite{ivanov2012nonmonotonic,subbotin1,subbotin2,zakinyan2020thermal}. 
Here, the non-uniformity of the internal magnetic field inside the sample leads to a very sophisticated magnetic response. 
It is, however, clear that the internal field gradients will become even stronger and more important, 
if a true binary magnetic material is addressed. 
So far, a detailed description as well as a fundamental understanding of the magnetisation processes in such materials is not available in the scientific literature. 
This work aims at filling this gap and puts forward a combined analytical-computational study of a system, 
containing both large superparamagnetic nanoclusters (the sources of strong internal field and spatial inhomogeneity) 
and a dispersed phase of single-domain superparamagnetic particles that are forced to react to the perturbations 
created by the cluster.
As long as we expect a drastic change depending on the carrier,
we investigate two extreme cases: the disperse phase is either frozen in space,
maintaining only the rotational degrees of freedom, mimicking a material, based on a rubber-like rigid matrix;
or the whole system is immersed in a liquid, where the disperse phase can freely diffuse. 
It turns out that in the latter, a pronounced gathering of a dispersed phase in the vicinity of the nanocluster is observed, causing qualitative changes in the magnetisation behaviour.

The article is organised as follows. First, we describe the model in detail in Sec.~\ref{sec:model}. 
In Sec.~\ref{sec:theory}, we adapt the analytical approach, developed by Subbotin~\cite{subbotin1,subbotin2}, and calculate the magnetization of our binary system. 
The results and discussions in Sec.~\ref{sec:results} are split, according to the carrier: we discuss solid matrix in~\ref{sec:solid}; 
a liquid carrier is studied in~\ref{sec:liquid} and~\ref{sec:halo} 
of the Results. 
In particular, spatial redistribution of a dispersed phase is investigated in Sec.~\ref{sec:halo}. 
The summary and a short outlook are provided in Sec.~\ref{sec:conc}.

\section{Model of a binary superparamagnetic mixture} \label{sec:model}

\begin{figure}
 \centering
 \includegraphics[scale=.25]{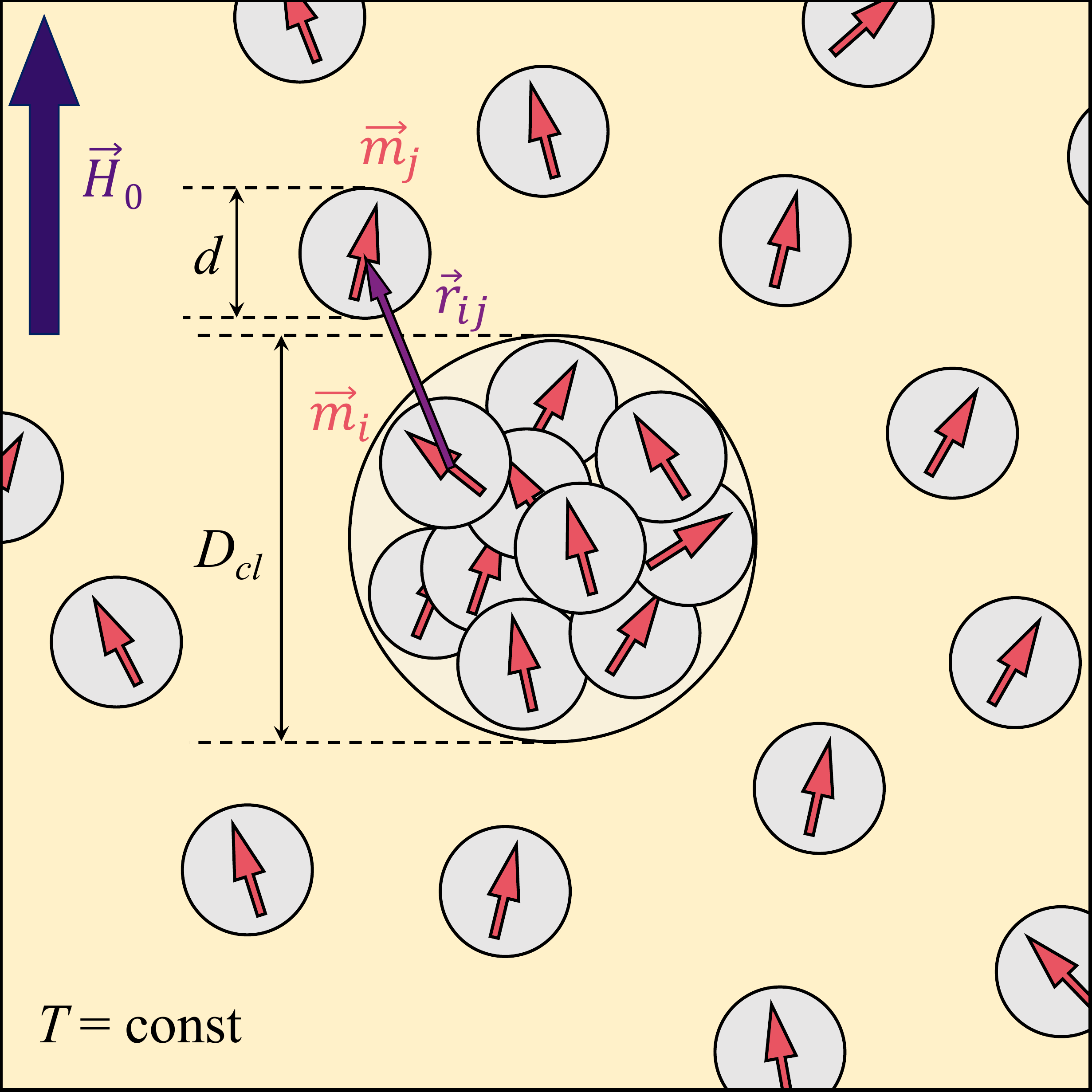}
 \caption{Schematic representation of the investigated system}
 \label{fig:1}
\end{figure}

The system under consideration is an isolated magnetic nanocluster embedded in a superparamagnetic medium (see Fig.~\ref{fig:1}).
The system is subjected to an external uniform magnetic field $\vec{H}_0$ and thermostated at a constant temperature $T$.
Nanocluster is modelled as a sphere of diameter $D_{cl}$
filled with $N_{in}$ spherical magnetic particles of diameter $d$. 
Particles are distributed within the cluster randomly and uniformly, 
without overlapping, their volume fraction is
\begin{equation} \label{eq:phi_in}
    \varphi_{in} = N_{in} \frac{v}{V_{cl}} = N_{in} \left(\frac{ d}{D_{cl}}\right)^3,
\end{equation}
where $v = (\pi/6)d^3$ and $V_{cl} = (\pi/6) D_{cl}^3$ 
are volumes of the particle and the nanocluster, respectively. 
Positions of particles within the nanocluster are rigidly fixed.  
Particles are assumed to be single-domain and magnetically-isotropic 
\hly{(the validity of this approximation is commented on in Appendix~\ref{sec:iso})}.
Each particle has a magnetic moment $\vec{m}$, which magnitude is fixed, but its
orientation can change under the influence of an applied magnetic field, 
dipolar magnetic fields created by other magnetic moments in the system and thermal fluctuations.
As a result, the nanocluster as a whole will exhibit a superparamagnetic behaviour according to the definition introduced in the previous section.
It does not have a net magnetic moment in the absence of an applied field,
but will be non-linearly magnetized, if the field is turned on~\cite{kuznetsov2018equilibrium}.
The superparamagnetic medium surrounding the nanocluster is modelled in a similar
fashion.
It consists of $N_{ex}$ magnetically-isotropic spherical single-domain particles,
which are exactly the same as particles that constitute the cluster,
{\it i.e.} they also have diameter $d$ and rotatable magnetic moment $\vec{m}$.
The particle volume fraction in the medium is 
\begin{equation} \label{eq:phi_ex}
    \varphi_{ex} = N_{ex} \frac{v}{V_{tot} - V_{cl}},
\end{equation}
where $V_{tot}$ is the total system volume.
Magnetic nanoparticles in the medium always retain their rotational degrees of freedom.

The interaction of magnetic moments with the external field is governed by the
Zeeman potential
\begin{equation}\label{eq:enh}
    U_Z = -\mu_0 (\vec{m} \cdot \vec{H}_0),
\end{equation}
where $\mu_0$ is the magnetic permeability of vacuum.
Additionally, each pair of particles interacts via 
the magnetic dipole-dipole potential
\begin{equation}\label{eq:endd}
U_{d d}(i,j)=\frac{\mu_0}{4 \pi} \left[ \frac{\left(\vec{m}_{i} \cdot \vec{m}_{j}\right)}{r^{3}_{ij}}-\frac{3\left(\vec{m}_{i} \cdot \vec{r}_{i j}\right)\left(\vec{m}_{j} \cdot \vec{r}_{i j}\right)}{r^{5}_{ij}}\right],
\end{equation}
where $\vec{m}_i$ and $\vec{m}_j$
are magnetic moments of two particles 
and $\vec{r}_{ij}$ is the vector connecting their centers.
We use two dimensionless energy parameters to characterize these magnetic interactions.
The first one is the Langevin parameter 
\begin{equation}
    \xi_0 = \frac{\mu_0 m H_0}{k_B T},
\end{equation}
that is the ratio of the Zeeman energy to the thermal energy $k_B T$,
$k_B$ is the Boltzmann constant, $m = |\vec{m}_i|$.
For a ten nanometer magnetite grain (with the saturation magnetization $M_s = 450~\textrm{kA/m}$),
$\xi_0 = 1$ corresponds to $H \simeq 14~\textrm{kA/m}$ at room temperature.
The second parameter is the so-called dipolar coupling constant
\begin{equation}
    \lambda = \frac{\mu_0}{4 \pi} \frac{m^2}{d^3 k_B T},
\end{equation}
that is the characteristic energy scale of two adjacent particles, whose dipoles are aligned head-to-tail, divided by $k_B T$ and calculated per particle.
At $T = 300~\textrm{K}$, this parameter for a pair of magnetite grains will roughly go from $\lambda \sim 1$ to $\lambda \sim 10$ as their diameter increases from 10 to 20 nanometers.

Modelling of different carriers will be done by changing the way
we treat the translational degrees of freedom of our particles.
In a solid carrier matrix (\textbf{SCM}), single-domain particles
surrounding the cluster will be randomly distributed in the medium 
and would not be able to move (just like the particles that constitute the nanocluster itself).
In a liquid carrier matrix (\textbf{LCM}), particles will be 
subjected to a translational Brownian motion
and could change their position relative to the nanocluster.
They are assumed to be sterically stabilized and are not allowed to overlap.
Of course, in a liquid the nanocluster should undergo the Brownian motion as well.
However, we can make use of the fact that characteristic time scales 
for 3D Brownian motion of a cluster and a particle,
$\tau_{cl} = 3 \eta V_{cl} / k_B T$ and $\tau_p = 3 \eta v / k_B T$, respectively 
($\eta$ is the carrier viscosity),
are very different.
Indeed, if $D_{cl}/d \sim 10$, the cluster motion is three orders of magnitude slower
than that of surrounding particles.
Thus, in LCM simulations the cluster will be treated as if its position is fixed. 

Our main quantities of interest in this work are the normalized 
equilibrium magnetic moment of the nanocluster
\begin{equation}
\vec{\cal{M}}_{cl} = \left\langle \sum^{N_{in}}_{i = 1} \vec{m}_i \right\rangle \frac{1}{m N_{in}},
\end{equation}
as well as the total normalized magnetic moment of the whole system
\begin{equation}
\vec{\cal{M}}_{tot} = \left\langle \sum^{N_{tot}}_{i = 1} \vec{m}_i \right\rangle \frac{1}{m N_{tot}},
\end{equation}
where $N_{tot} = N_{in} + N_{ex}$ is the total number of particles in the system,
$\langle \ldots \rangle$ denotes an ensemble average.

In this work, we use both analytical theory (Section~\ref{sec:theory}) and 
Langevin dynamics simulations (for details, see Appendix~\ref{sec:sim}). However, already at this point, it is important to specify that the system in simulations is 
subjected to 3D periodic boundary conditions.
It approximately corresponds to a suspension of nanoclusters
with a nanocluster volume fraction
\begin{equation}
    \Phi_{cl} = \frac{V_{cl}}{V_{tot}} = \left(1 + \frac{\varphi_{in}}{\varphi_{ex}}\frac{N_{ex}}{N_{in}}\right)^{-1}.
\end{equation}
We will consider systems with $N_{in} = 500$ and $N_{ex} = 2500$.
Particle concentration in the cluster is always $\varphi_{in} = 0.3$ 
(correspondingly, $D_{cl} \simeq 12d$),
while the concentration of the surrounding medium will be changed
from a small value of $\varphi_{ex} = 0.002$ to $\varphi_{ex} = 0.15$.
Correspondingly, the cluster concentration will change from $\Phi_{cl} \simeq 0.0013$
to $\Phi_{cl} \simeq 0.09$.
Magnetic interaction parameters also will vary in wide ranges:
$1 \le \lambda \le 5$, $0 \le \xi_0 \le 5$.

\section{Magnetic response theory} \label{sec:theory}

In this section, we will summarize the works of Subbotin on inverse ferroemulsions~\cite{subbotin1,subbotin2} 
and adapt them to binary superparamagnetic mixtures.
Let us consider a suspension of spherical magnetizable bodies (clusters) 
in a magnetizable medium with a relative magnetic permeability $\mu_{ex}$.
Assume that the volume fraction of clusters is $\Phi_{cl}$ and they are made of some material 
with relative magnetic permeability $\mu_{in}$.
According to Refs.~\cite{subbotin1,subbotin2}, the field
inside clusters is homogeneous and parallel to the external field,
its magnitude is given by
\begin{equation} \label{eq:hin}
    H_{in} = H_0 \frac{1}{1 + (1-\Phi_{cl})\kappa\left( \frac{\mu_{in}}{\mu_{ex} }- 1\right)},
\end{equation}
where $\kappa$ is the cluster demagnetization factor.
For a sphere, $\kappa~=~1/3$.
The field in the surrounding medium is
\begin{equation} \label{eq:hex}
    H_{ex} = H_0 \left[ 1 + \frac{\Phi_{cl} \kappa \left(\frac{\mu_{in}}{\mu_{ex}} - 1\right)}{1 + (1-\Phi_{cl})\kappa\left( \frac{\mu_{in}}{\mu_{ex} }- 1\right)} \right].
\end{equation}
Permeabilities, in general, can be considered as non-linear functions of the field:
\begin{gather}
    \mu_{in} = 1 + \frac{M_{in}(H_{in})}{H_{in}}, \label{eq:muin}\\
    \mu_{ex} = 1 + \frac{M_{ex}(H_{ex})}{H_{ex}}, \label{eq:muex}
\end{gather}
where $M_{in}$ and $M_{ex}$ are magnetizations of the cluster material and of the medium,
respectively. 
The total magnetization of the system is
\begin{equation} \label{eq:magn_tot}
    M_{tot} = \Phi_{cl} M_{in}(H_{in}) + (1 - \Phi_{cl}) M_{ex}(H_{ex}). 
\end{equation}
Normalized magnetic moments then can be found
as 
\begin{gather}
    \mathcal{M}_{cl} = \frac{M_{in}V_{cl}}{m N_{in}}, \:\:\: 0 \le \mathcal{M}_{cl} \le 1, \label{eq:mm_cl}\\
    \mathcal{M}_{tot} = \frac{M_{tot}V_{tot}}{m N_{tot}}, \:\:\: 0 \le \mathcal{M}_{tot} \le 1 \label{eq:mm_tot}.
\end{gather}
Further on, the set of equations~(\ref{eq:hin})--(\ref{eq:mm_tot}) will be referred to 
as the {binary mixture magnetization} (\textbf{BMM}) model.

The key assumption of BMM is that magnetic field in the medium $H_{ex}$ 
is a sum of the external field $H_0$ and some average field that is created by all magnetized clusters, distributed in the system.
This latter field is assumed to be uniform and so is $H_{ex}$
itself.
However, it is known that the local magnetic field (and subsequently $\mu_{ex}$) 
in the vicinity of a magnetized spherical body is non-uniform~\cite{rosensweig1985ferrohydrodynamics}.
Thus, Eqs.~(\ref{eq:hin})--(\ref{eq:hex}) are only an approximation.
BMM, however, converges to a well-known Maxwell-Wagner formula for the initial permeability
of a binary dielectric mixture~\cite{fricke1953maxwell}.
In the weak-field limit, it also shows a good agreement with experimental data on the effective permeability of inverse ferroemulsions. 
This model, however, overestimates experimental results slightly as the applied field increases.
Applicability of BMM to our system is to be determined.

In order to close the set of BMM equations, some explicit expressions 
for magnetization curves $M_{in} = M_{in}(H_{in})$ and $M_{ex} = M_{ex}(H_{ex})$ are required.
For this purpose, the so-called {modified mean-field} (\textbf{MMF})
theory can be used.
It was initially developed to describe static magnetic properties 
of concentrated ferrofluids~\cite{pshenichnikov1996magneto,ivanov2001magnetic}.
Subsequently, the approach has been 
extended for the description of the ferrofluid dynamic response~\cite{ivanov2016revealing}
as well as magnetic properties of single-domain nanoparticle
ensembles immobilized in a solid non-magnetic matrix~\cite{elfimova2019static}.
Ref.~\cite{subbotin2} also used the first-order MMF
to describe the magnetic component of an inverse ferroemulsion.
Within a more accurate second-order MMF approach~\cite{ivanov2006magnetogranulometric},
magnetization of a one-component superparamagnetic material can be written as
\begin{gather}
    M(H) = M_s L\Biggl(\xi_{eff}\left(\frac{\mu_0 m H}{k_B T}, \chi_L\right)\Biggl), \label{eq:mmf1}\\
    \xi_{eff}(\xi, \chi_L) = \xi + \chi_L \left(1 + \chi_L L'(\xi)/16 \right) L(\xi),\label{eq:mmf2}\\
    L(\xi) = \coth \xi - 1/\xi,
\end{gather}
where $M_s = (6/\pi d^3) m \varphi$ is the material saturation magnetization, 
$\varphi$ is the particle volume fraction,
$\chi_L = 8 \lambda \varphi$ is the so-called Langevin susceptibility,
$L(\xi)$ is the Langevin function that describes magnetic response of an ensemble of non-interacting dipoles ({\it i.e.}, at $\chi_L \ll 1$),
$L'(\xi) = d L(\xi)/d \xi$, 
$\xi_{eff}$ is the effective dimensionless field that is acting locally on an arbitrary chosen 
particle in an ensemble with dipole-dipole interactions.
If we assume that both components of our binary mixture can be described by MMF, 
permeabilities can be written down as
\begin{gather}
    \mu_{in} = 1 + 3 \chi^{in}_L \frac{L\left(\xi_{eff}(\xi_{in},\chi^{in}_L)\right)}{\xi_{in}}, \label{eq:muin_mmf}\\
    \mu_{ex} = 1 + 3 \chi^{ex}_L \frac{L\left(\xi_{eff}(\xi_{ex},\chi^{ex}_L)\right)}{\xi_{ex}},\label{eq:muex_mmf}
\end{gather}
where $\chi^{in}_L = 8 \lambda \varphi_{in}$, $\chi^{ex}_L = 8 \lambda \varphi_{ex}$,
$\xi_{in} = \mu_0 m H_{in}/k_B T$, $\xi_{ex}~=~\mu_0 m H_{ex}/k_B T$.

\begin{figure*}[t]
 \centering
 \includegraphics[scale=0.85]{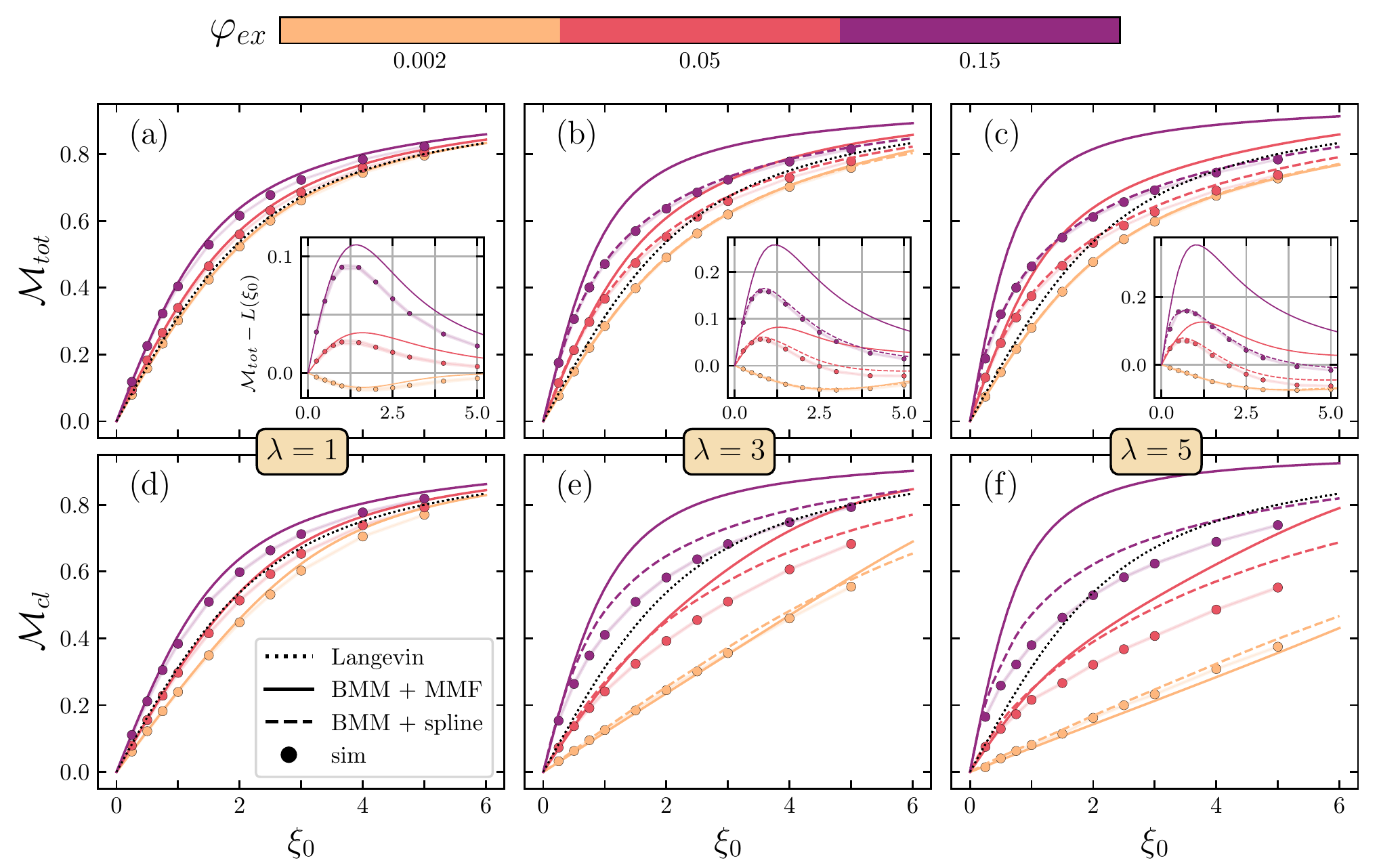}
 \caption{Equilibrium magnetization curves of a superparamagnetic mixture in a solid carrier. The first row (panels (a)--(c)) demonstrates dependencies of a normalized magnetization (or, identically, of a normalized magnetic moment) of the whole system $\mathcal{M}_{tot}$ on the Langevin parameter $\xi_0$. Insets 
 in (a)--(c) shows the difference between $\mathcal{M}_{tot}$ values from the corresponding panels and the Langevin function $L(\xi_0)$. 
 The latter is indicated on the main panels with \textbf{dotted lines}.
 The second row (panels (d)--(f)) shows corresponding values of the cluster normalized magnetic moment $\mathcal{M}_{cl}$. Different columns correspond to different dipolar coupling parameters: (a), (d) $\lambda = 1$; (b),~(e)~$\lambda = 3$; (c),~(f)~$\lambda = 5$. 
 Particle volume fraction in the surrounding medium $\varphi_{ex}$
 is indicated by color (see colorbar). 
 Simulation results are shown with \textbf{circles} 
 (transparent lines connecting them are guides for eyes),
 \textbf{solid lines} are predictions from BMM model [Eqs.~(\ref{eq:hin})-(\ref{eq:mm_tot})] combined with MMF expressions for magnetic permeabilities [Eqs.~(\ref{eq:muin_mmf})-(\ref{eq:muex_mmf})].
 \textbf{Dashed lines} are ``corrected'' BMM predictions with permeabilities 
 values directly extracted from auxiliary simulations of one-component superparamagnetic systems rather than from MMF.
 }
 \label{fig:2}
\end{figure*}

\begin{figure}
 \centering
 \includegraphics[scale=.8]{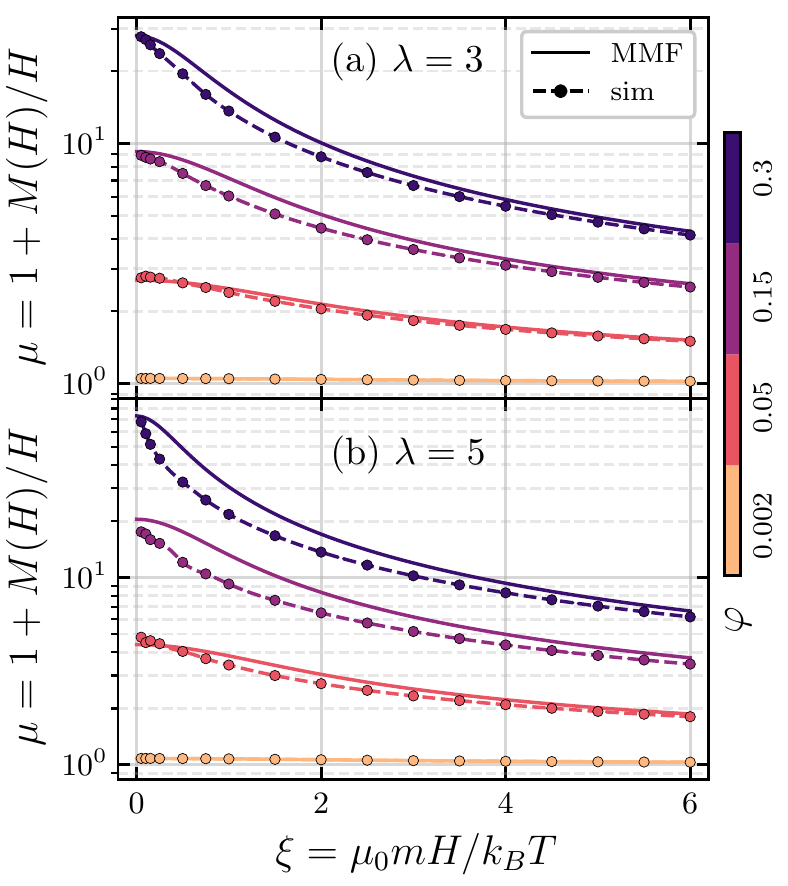}
 \caption{Field dependencies of the magnetic permeability for a one-component ensemble of randomly distributed immobilized magnetic nanoparticles. 
 \textbf{Solid lines} are MMF theory predictions [Eqs.~(\ref{eq:mmf1})-(\ref{eq:mmf2})], \textbf{symbols} are simulation results.
 Different panels correspond to different dipolar coupling constants: $\lambda = 3$~(a) and 5~(b). Particle volume fractions are indicated by color.}
 \label{fig:3}
\end{figure}

\section{Results and discussion}\label{sec:results}
\subsection{Equilibrium magnetization of a mixture in a solid carrier} \label{sec:solid}

Magnetization curves for a superparamagnetic cluster embedded in a solid matrix with immobilized nanoparticles are shown in Fig.~\ref{fig:2} for different values of $\lambda$ and $\varphi_{ex}$. 
The first noticeable feature of magnetization curves
is that at any given $\lambda$ an increase in $\varphi_{ex}$
leads to a qualitative change in how the system magnetization relates to the Langevin function.
Langevin magnetization corresponds to a system of non-interacting dipoles. So, if the normalized magnetization is lower than corresponding Langevin value,
it means that dipole-dipole interactions hinder the overall magnetic response. 
{\it Vice versa}, magnetization higher than the Langevin value indicates that dipole-dipole interactions play a reinforcing role. 
It is known that equilibrium magnetostatic response of superparamagnetic clusters in an empty space always lies below the Langevin curve~\cite{kuznetsov2018equilibrium} --
this is the result of the demagnetization effect.
Similar behavior is observed in our system for a cluster in a diluted medium with $\varphi_{ex} = 0.002$.
However, as the concentration of particles in the surrounding medium increases (and as the magnetic permeability of the medium $\mu_{ex}$ becomes closer to the permeability of the cluster $\mu_{in}$),
demagnetization effects slowly disappear -- magnetization of both the cluster and the mixture eventually becomes larger than the Langevin value.
Interestingly enough, at  $\varphi_{ex} \ge 0.05$
and $\lambda \ge 3$ the impact of dipole-dipole interactions on the mixture magnetization depends non-monotonically on the field -- while the initial section of the magnetization curve is larger than that of the Langevin function, simulation points eventually fall below $L(\xi)$ in the saturation regime.    

As for the theoretical predictions, it is seen
that the combination of BMM and MFT gives very accurate predictions for the initial slope of magnetization curves in the whole investigated parameter ranges.  
However, as the field increases, theory and simulation data start to diverge rapidly.
The larger $\varphi_{ex}$ and/or $\lambda$ 
reinforce the discrepancy. 
Theoretical (solid) curves  in Fig.~\ref{fig:2} 
are always above simulation points
at $\xi_0 > 1$.
At $\varphi_{ex} = 0.15$ and $\lambda = 5$,
the error between numerical and theoretical values of the mixture magnetization reaches almost 20\% of the corresponding saturation value.

To understand the reason for the discrepancy between theory and simulations, 
a set of auxiliary simulations was performed.
We simulated a one-component ensemble of immobilized nanoparticles randomly and uniformly distributed in a standard cubic box with 3D periodic boundary conditions. Ensemble of $N = 2000$ particles 
was considered.
Using simulation data, a non-linear magnetic permeability
of the ensemble was calculated as a function of the field $\xi$
at different $\lambda$ and particle volume fractions~$\varphi$.
The results are demonstrated in Fig.~\ref{fig:3}
in comparison with MMF predictions.
It can be seen that while MMF mostly predicts correct zero-field
permeability values, at large fields it overestimates $\mu$.
In more details this feature of immobilized superparamagnetic ensembles was discussed in Ref.~\cite{kuznetsov2018equilibrium}.
To take it into account the following procedure was performed.
Calculated permeability of a one-component system were interpolated (with cubic splines)
and then put in BMM instead of MMF predictions Eqs.~(\ref{eq:muin_mmf}) and (\ref{eq:muex_mmf}).
The results of this procedure are shown in Fig.~\ref{fig:2} with 
dashed lines.
We can see that the accuracy of BMM with ``corrected'' permeabilities
improves drastically. 
New theoretical curves closely follow $\mathcal{M}_{tot}$ dependencies.
The magnetization curves for the cluster still overestimate 
numerical results but it is much better than MMF for $\lambda \ge 3$
and $\varphi_{ex} \ge 0.05$.
The probable reason for the remaining discrepancy is the inherent BMM assumption
that the magnetic field $H_{ex}$ and permeability of the surrounding medium $\mu_{ex}$ are constant and uniform  in cluster vicinity, which is not correct at large enough applied fields~\cite{rosensweig1985ferrohydrodynamics}. 
Thus, the agreement with simulation potentially can only 
be improved by directly solving a nonlinear magnetostatic boundary-value problem  
and correctly determining the magnetic field distribution in the system.
However, this task is beyond the scope of the present paper.

\begin{figure*}[t!]
 \centering
 \includegraphics[scale=0.85]{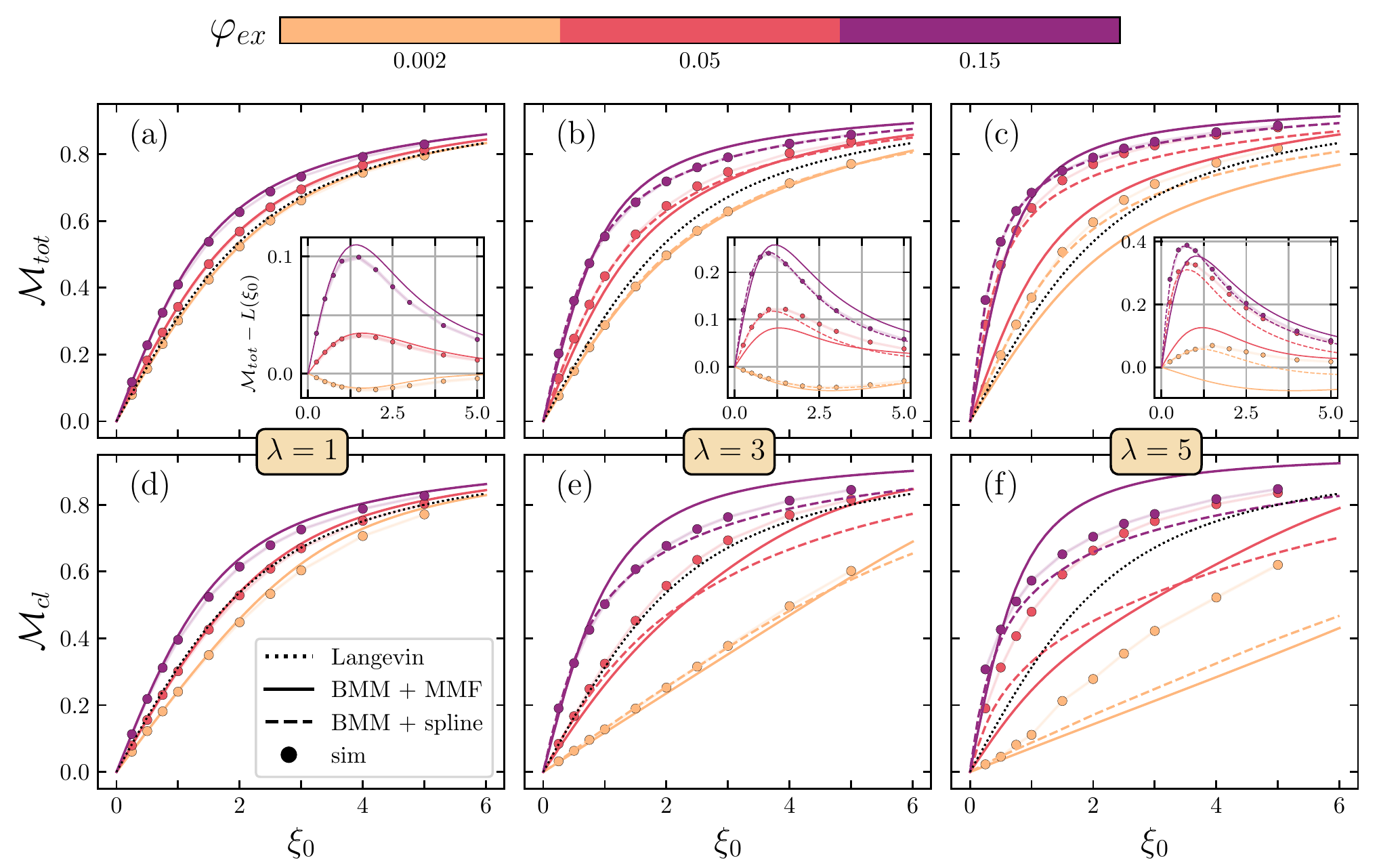}
 \caption{Equilibrium magnetization curves for a superparamagnetic mixture in a liquid carrier. 
 The notation is identical to Fig.~\ref{fig:2}. 
 Note that colored solid curves corresponding to MMF predictions 
 are also exactly the same as in Fig.~\ref{fig:2}. 
 }
 \label{fig:4}
\end{figure*}

\begin{figure}[t!]
 \centering
 \includegraphics[scale=.8]{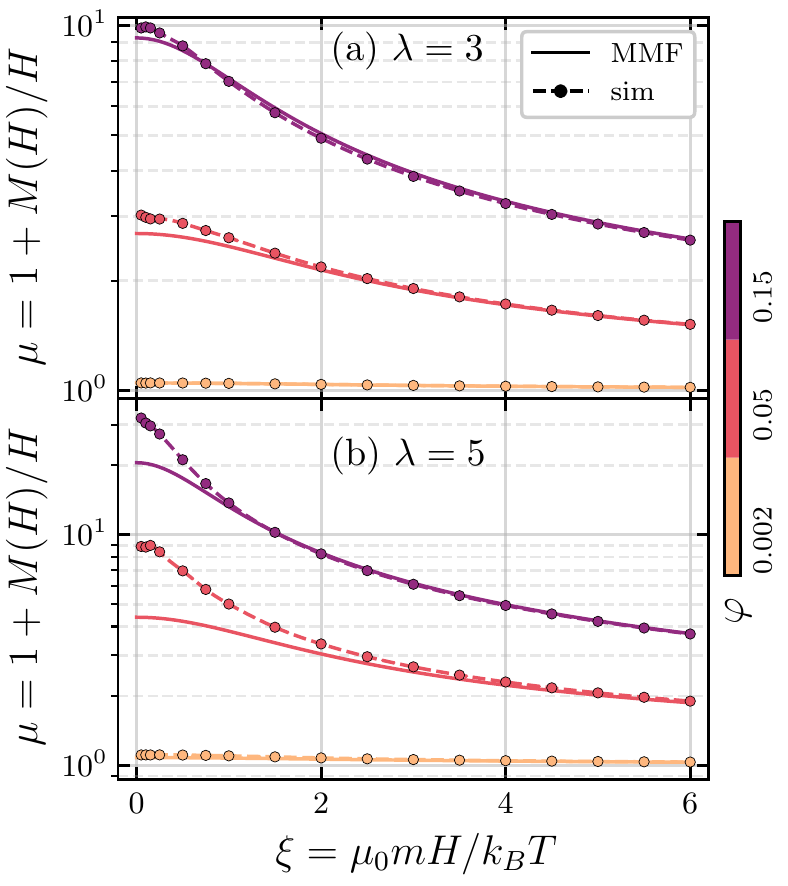}
 \caption{Field dependencies of the magnetic permeability for a one-component ensemble of magnetic nanoparticles suspended in a liquid matrix (i.e., particles are subjected to a translational Brownian motion). The notation is identical to Fig.~\ref{fig:3}. }
 \label{fig:5}
\end{figure}

\subsection{Equilibrium magnetization of a mixture in a liquid carrier} \label{sec:liquid}

Now let us consider a different type of a binary mixture --
a superparamagnetic nanocluster submerged in 
a suspension of magnetic nanoparticles in a non-magnetic liquid matrix.
Essentially, a nanocluster in a ferrofluid or in a very loose gel, in which the cluster is too large to diffuse, 
but a dispersed phase  is not constrained \cite{krekhova2010a}.
The magnetization curves for this case are shown in Fig.~\ref{fig:4}.
The first thing that is seen here is that the magnetization of the mixture and the cluster are larger than corresponding SCM values
for every set of investigated parameters.
A noticeable feature of previously considered SCM is that  $\mathcal{M}_{tot}$ at large $\varphi_{ex}$ can be higher than the corresponding Langevin value in weak fields, but smaller than the Langevin value in strong fields. 
The role of dipole-dipole interactions changes as the field increases.
For LCM, this feature is no longer present -- there are no crossing of the Langevin  curve,
at least not at $\xi_0 \le 5$.

MMF does not make any distinctions between liquid and solid superparamagnetic ensembles -- thus, theoretical curves in Fig.~\ref{fig:4} 
are exactly the same as in Fig.~\ref{fig:2}.  
At $\lambda = 1$, these curves actually
describe simulation data for LCM 
quite well -- better than the data for SCM (compare insets in Fig.~\ref{fig:2}(a) and Fig.~\ref{fig:4}(a)). 
But already at $\lambda = 3$ the agreement breaks down.
Surprisingly, the error does not increase with $\varphi_{ex}$ as in SCM case -- the strongest disagreement between theory and simulations takes place at intermediate and low concentrations.
For $\lambda = 3$ and $\varphi_{ex} = 0.05$, theoretical predictions are incorrect both for initial and saturation portions of the LCM magnetization curve (Figs.~\ref{fig:4}(b),(e)).
At $\lambda = 5$, the strongest disagreement takes place at even smaller concentrations,
$\varphi_{ex} = 0.002$ (Figs.~\ref{fig:4}(c),(f)).

In order to improve the agreement, the same procedure was employed as
for SCM. 
Namely, an auxiliary set of simulations of a one-component superparamagnetic system was performed. 
This time, the one-component system was a liquid suspension of single-domain particles.
The results for a non-linear magnetic permeability of this system at different particle concentrations and dipolar coupling constants are given in Fig.~\ref{fig:5}.
Comparing it to Fig.~\ref{fig:3}, we can see that the relation between actual permeability and MMF predictions for liquid and solid one-component superparamagnets is completely opposite.
\hly{For solid systems, zero-field permeabilities are correctly described by MMF, 
but the theory overestimates the magnetic response as the field increases.
These features of solid superparamagnets are well documented in 
the literature~\cite{kuznetsov2018equilibrium,elfimova2021pre}.
For a liquid, zero-field permeabilities are larger than MMF 
predictions, but in strong fields the agreement significantly improves.
This behavior can be attributed to the particle self-assembly,
which is not taken into account within the MMF framework.
It is known that magnetic particles with sufficiently strong dipole-dipole interactions 
tend to form chain-like aggregates in viscous~\cite{wang2002molecular} 
and even in soft elastic environments~\cite{pessot2018}.   
In monodisperse ferrofluids, the chain formation increases the initial magnetic response~\cite{ivanov2004applying},
but under saturation condition the influence of chains on the magnetization reduces~\cite{mendelev2004ferrofluid}.}

Once again, numerically obtained permeability curves were interpolated with cubic splines and then used within BMM approach instead of MMF predictions Eqs.~(\ref{eq:muin_mmf}) and (\ref{eq:muex_mmf}).
The results of this correction are shown in Fig.~\ref{fig:4} with
dashed lines.
Unfortunately, the correction no longer gives the same accuracy boost as for SCM.
In fact, it only improves the initial slope of the magnetization curves.
But at large fields, simulation results persistently
lie above the predictions of the corrected BMM. 
It is most clearly demonstrated by $\mathcal{M}_{tot}$ and $\mathcal{M}_{cl}$
dependencies for $\lambda = 5$ and $\varphi_{ex} = 0.002$ (Fig.~\ref{fig:4}(c), (f)).

So, it can be deduced that some qualitative change is happening in LCM
system, as the field increases:
\begin{itemize}
    \item it leads to a significant increase of normalized magnetic moments $\mathcal{M}_{tot}$ and $\mathcal{M}_{cl}$;
    \item it cannot be explained within BMM approach;
    \item it is more pronounced at larger $\lambda$ and smaller $\varphi_{ex}$;
    \item it does not take place in SCM.
\end{itemize}

\begin{figure*} 
 \centering
 \includegraphics[scale=.85]{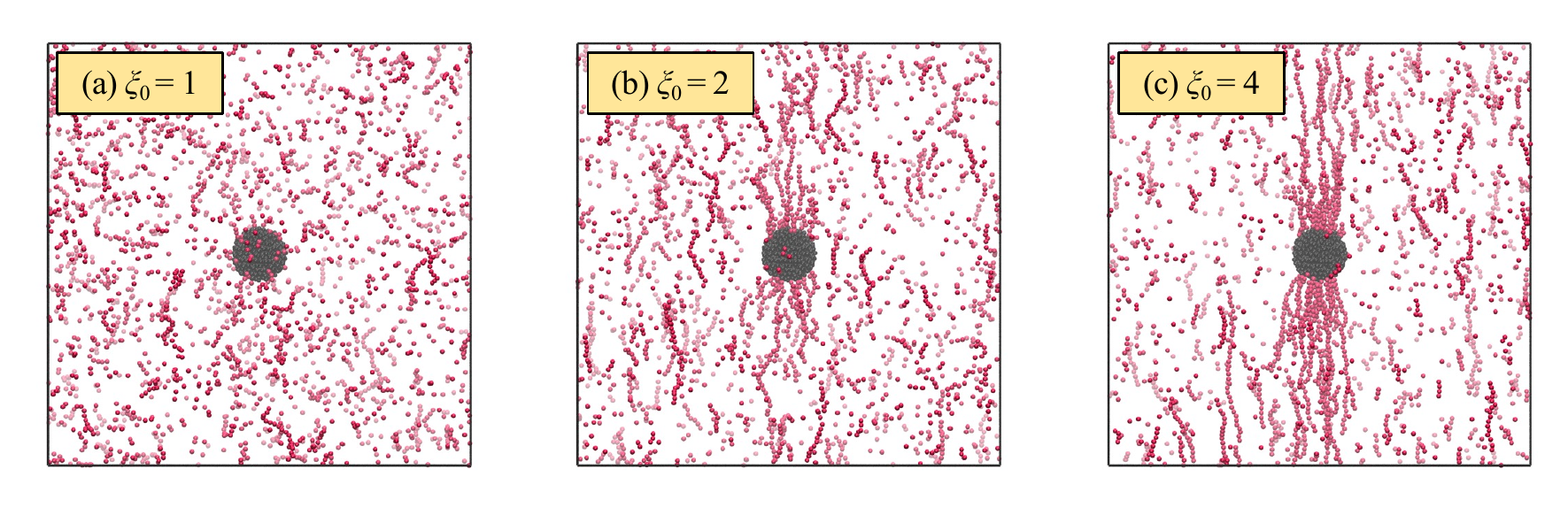}
 \caption{Simulation snapshots of the system in a liquid carrier at $\varphi_{ex} = 0.002$ and $\lambda = 5$. Different panels correspond to different Langevin parameters: $\xi_0 = 1$~(a), 2 (b) and 4 (c). Applied field is oriented vertically.}
 \label{fig:6}
\end{figure*}

To get a better understanding of what is happening here,
a deeper analysis of the system microstructure is presented below.

\begin{figure} 
 \centering
 \includegraphics[scale=.57]{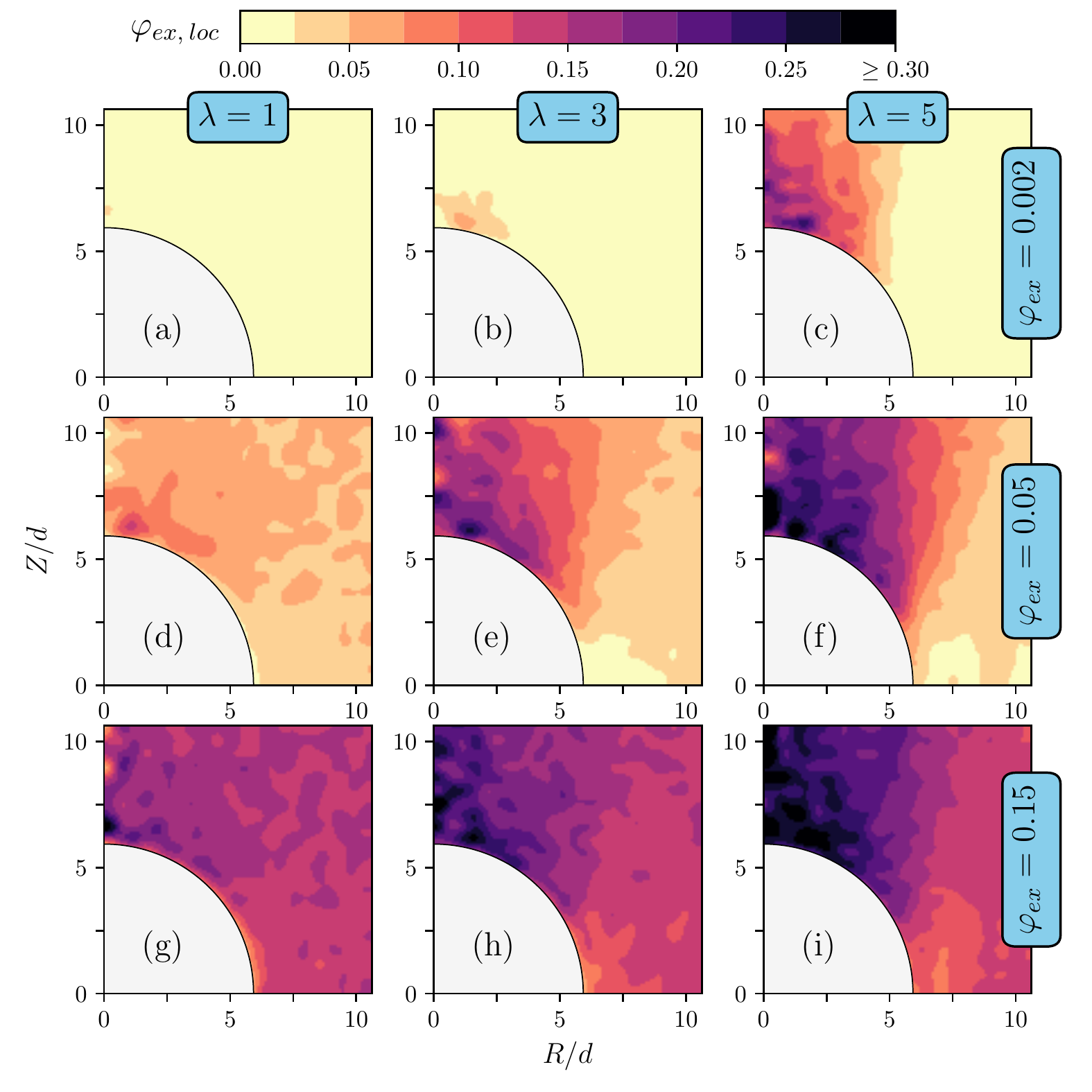}
 \caption{\textit{Local} particle volume fraction $\varphi_{ex,\:loc}$ in the vicinity of the cluster at $\xi_0 = 3$. 
 Numerical values of $\varphi_{ex,\:loc}$ are indicated by the color (see colorbar),
 the cluster itself is colored grey.
 Maps are constructed using space- and time-averaged data from 3D Langevin dynamics simulations.
 They are plotted in a cylindrical coordinates ($R, Z$) with the origin at the cluster center.
 Only the upper right corner is shown due to the system symmetry.
 The field is directed along $Z$ axis.
 Dipolar coupling constant is increasing from left to right: 
 (a),(d),(g) $\lambda = 1$;
 (b),(e),(h) $\lambda = 3$; 
 (c),(f),(i) $\lambda = 5$.
 The \textit{average} volume fraction of particles is increasing from top to bottom:
 (a)--(c) $\varphi_{ex} = 0.002$;
 (d)--(f) $\varphi_{ex} = 0.05$;
 (g)--(i) $\varphi_{ex} = 0.15$.
 }
 \label{fig:7}
\end{figure}

\subsection{Field-controlled haloing in a liquid carrier} \label{sec:halo}

Let us look closely on the behavior of the simulated LCM system 
at $\lambda = 5$ and $\varphi_{ex} = 0.002$ -- \textit{i.e.}, in the parameter ranges, 
where the deviations from theoretical magnetization curves are most pronounced.
Corresponding simulation snapshots are collected 
in Fig.~\ref{fig:6} for different values of the applied field strength.  
At a relatively small field, $\xi_0 = 1$,
free nanoparticles form chain-like structures,
as is expected at $\lambda = 5$~\cite{ivanov2004applying}. 
The presence of a cluster does not produce any clearly visible effects on the system microstructure.
However, already at $\xi_0 = 2$ and, especially, 
at $\xi_0 = 4$ a significant change takes place --
particles (or rather, particle chains) 
start to concentrate near the nanocluster poles forming 
clouds stretched along the field direction. 

The described phenomenon is qualitatively reminiscent of the so-called ``\textit{haloing}'' effect, experimentally observed in bimodal magnetorheological fluids~\cite{magnet2012haloing,magnet2014behavior}.
The only difference is that in the latter systems 
superparamagnetic nanoclusters form thick clouds (or ``\textit{halos}'') around a magnetizable microsphere.
On an even larger scale the phenomenon was reproduced in Ref.~\cite{ivanov2014vortex}:
authors observed the condensation of drop-like aggregates 
of a phase-separated ferrofluid on the surface
of a millimeter-sized iron sphere.
The physical reason behind this halo formation in both cases is the phenomenon of magnetophoresis -- \textit{i.e.},
the motion of magnetic nanoparticles in a gradient magnetic field~\cite{pshenichnikov2012magnetophoresis, kuznetsov2021magnetophoretic}.
The source of the inhomogeneous field in our problem is the magnetized nanocluster~\cite{rosensweig1985ferrohydrodynamics}. 
The stronger the applied field, the stronger the cluster's own field gradient. 
This gradient is directed towards poles of the cluster, where free nanoparticles and nanoparticle chains tend to accumulate.

It is known that the transport of magnetic nanoparticles 
in a viscous medium is affected strongly by interparticle interactions~\cite{pshenichnikov2011magnetophoresis,kuznetsov2017sedimentation}.
Namely, dipole-dipole interactions, controlled by $\lambda$, 
are acting as effective attraction between particles.
They decrease the gradient diffusion coefficient of the system 
and make it much easier to create a highly inhomogeneous particle distribution with a given applied field. 
The effect of dipole-dipole interactions on particle transport is typically most pronounced at intermediate average concentrations $\varphi \leq 0.1$.
In more dense systems, the steric repulsion (\textit{i.e.}, the excluded volume effect) starts to dominate 
and substantially increases the gradient diffusion coefficient. 
To put it simply, it is hard to create a noticeable concentration gradient in a highly concentrated system. 
All these theoretical considerations are well illustrated and validated by LCM concentration maps shown in Fig.~\ref{fig:7}. 
First, halos concentration increases with $\lambda$.
At $\varphi_{ex} = 0.002$ and $\lambda = 5$,
the \textit{local} concentration of particles near cluster poles is actually comparable with the cluster concentration itself ($\varphi_{in} = 0.3$) and two order of magnitude higher than near the cluster ``flanks''.
Thus, the situation can be interpreted as follows --
the cluster, which is spherical in small fields, 
start to absorb free particles with increasing $\xi_0$ 
and turns into an elongated aggregate aligned with the field.
As the aggregate shape changes, its demagnetization factor [$\kappa$ in Eqs.~(\ref{eq:hin})-(\ref{eq:hex})]
decreases. 
It is very similar to the behavior of magnetic droplets in ferroemulsions~\cite{ivanov2012nonmonotonic},
and can explain the anomalous increase of the cluster magnetization seen in Fig.~\ref{fig:4}(f).
The haloing is still present at larger average concentrations.
However, an important difference is that at higher average volume fractions
the inhomogeneity of the local concentration decreases.
In the most dense environment with the average particle concentration $\varphi_{ex} = 0.15$, 
the local particle concentration near cluster surface is always $\varphi_{ex,\:loc} \ge 0.1$.
Correspondingly, variations of $\mu_{ex}$ in the cluster vicinity are getting smaller.
As the result, BMM (which assumes the system homogeneity) works much better for concentrated LCM samples.  

\section{Conclusions} \label{sec:conc}

In this work, equilibrium magnetic response 
of a binary superparamgnetic mixture 
is studied both theoretically and numerically (with the help of Langevin dynamic simulations).
One component of the mixture is a spherical nanocluster, 
consisting of immobilized magnetically-isotropic single-domain particles.
The cluster is submerged in a superparamagnetic medium, 
which itself constitutes an ensemble of single-domain particles in a non-magnetic matrix.
Two cases are separately considered.
In the first case (SCM), particles of the surrounding medium are spatially immobilized, 
although they fully retain rotational degrees of freedom.
This case allows us to neglect possible effects of Brownian motion and particle aggregation.
In the second case (LCM), particles in the medium have both translation and rotational degrees of freedom.   
It is shown that magnetostatic response 
of the SCM system can be accurately described theoretically within BMM approach [Eqs.~(\ref{eq:hin})-(\ref{eq:mm_tot})],
if the non-linear permeabilities of individual mixture components are known.
It is also shown that MMF predictions for
permeabilities [Eqs.~(\ref{eq:muin_mmf})-(\ref{eq:muex_mmf})] give accurate description of the simulation data 
only at relatively small values of the dipolar coupling constant ($\lambda \leq 1$).
The situation changes qualitatively for LCM system. 
If the average particle concentration in the medium is low enough, 
magnetization of the mixture grows anomalously fast with the field (compared to BMM prediction).
The apparent reason for this growth is the so-called haloing effect: 
the gradient field of the magnetized nanocluster leads to 
the local redistribution of particles in the surrounding medium. 
It is shown that particles form concentrated clouds near the cluster poles, 
this way effectively reducing the demagnetization effect 
and making it more susceptible to the applied field.  
A strong dependence of the haloing effect on the intensity of dipole-dipole interactions is revealed.

We can conclude that an accurate theoretical description of the magnetostatic response 
of a binary superparamagnetic mixture at $\lambda > 1$ cannot simply assume the spatial homogeneity
of the systems magnetic properties.  
The local inhomogeneity of the cluster field and the subsequent drift-diffusion particle transport 
must be explicitly taken into account. 
In practice, it will require the solution of a combined magneto-diffusive boundary value problem, 
similar to those previously considered in Refs.~\cite{lavrova2016modeling,lavrova2022numerical}.
The solution of this problem is left for future studies.

\section*{Conflicts of interest}
There are no conflicts to declare.

\section*{Acknowledgements}
The study was funded by RFBR, project number 19-31-60036.
All computations were performed at the Ural Federal University cluster.
S.S.K. acknowledges the support from FWF Project SAM P 33748.

\appendix

\section{\hly{Approximation of magnetically-isotropic nanoparticles}} \label{sec:iso}

Let us give here a more detailed explanation of our particles being ``magnetically-isotropic''.

The simplest and most common model for an internal magnetic anisotropy of 
single-domain particles is the easy axis anisotropy~\cite{coffey2012thermal}.
It assumes that the particle possesses a special internal direction (easy axis),
which can be characterized by a unit vector $\hat{n}$.
The orientational coupling between the axis and the magnetic moment of any given particle 
is described by the potential

\begin{equation}
    U_a = - K v (\hat{m}\cdot\hat{n})^2,
\end{equation}

\noindent where $K$ is the particle anisotropy constant,
$\hat{m} = \vec{m}/m$.
The interplay between anisotropy and thermal fluctuations is
described by the dimensionless anisotropy parameter

\begin{equation}
    \sigma = \frac{Kv}{k_B T}.
\end{equation}

\noindent For a 10 nm particle with $K \sim 10^4~\textrm{J/m}^3$, 
the anisotropy parameter is $\sigma \sim 1$.

It is known, that the variation in $\sigma$ strongly affects the \textit{dynamic}
magnetic response of single-domain particles in different matrices~\cite{poperechny2010dynamic,ilg2022prb}.
However, in this work we are interested only in the equilibrium magnetic response to a \textit{static} applied field.
And it is known that equilibrium magnetization curves of superparamagnetic particles
suspended in a liquid simply do not depend on~$\sigma$~\cite{elfimova2019static}.
Situation becomes more complicated, if particles are immobilized in a solid carrier.
In principle, now one has to take into account the ``magnetic texture'' of the system,
\textit{i.e.}, the specific orientational distribution of particles' easy axes.
The texture can be created by applying a strong field during the synthesis stage
and can have a major impact on the system magnetic response~\cite{radushnov2022structure,raikher1983magnetization}.
However, non-textured composites with random and uniform distribution of easy axes
are more similar to liquid superparamagnets -- 
their initial magnetic response also does not depend on the anisotropy parameter $\sigma$~\cite{schaller2009effective,elfimova2019static}.
Increasing $\sigma$ can lower magnetic response at larger fields (at $\xi_0 > 2$),
but it will not affect the magnetization curve qualitatively~\cite{kuznetsov2018equilibrium}.
Taking all said into consideration, we decided to investigate here only 
the limiting case $\sigma \ll 1$, \textit{i.e.}, 
in zero field all internal orientations of 
magnetic moments are equiprobable.
Such approach allows us to considerably simplify both the theoretical treatment 
and the simulation protocol.
We believe that all the results obtained here for magnetically-isotropic systems 
can be extrapolated to liquids and non-textured solids with finite~$\sigma$.

\section{Simulation details} \label{sec:sim}

In a numerical realization of our mixture model, 
we consider a cubic simulation box with length 
$l = V_{tot}^{1/3}$.
The cluster is placed in the center of this box.
Its position does not change during the simulation.
3D periodic boundary conditions are imposed on a box.
The field is directed along $Z$-axis.
All the results reported are obtained using ESPResSo 4.1.4 simulation package~\cite{weik2019espresso}.

Rotational motion of the $i$-th particle is governed by the Langevin equation
\begin{equation}
    J^* \frac{d \vec{\omega}^*_i}{d t^*} = \vec{\tau}^*_i - \gamma^{*R}  \vec{\omega}^*_i + \vec{\eta}^{*R}_i, \:\:\: \frac{d \vec{m}_i }{dt^*} = \vec{\omega}^*_i \times \vec{m}_i. \label{eq:lang_rot}
\end{equation}
For LCM, translational motion of the $i$-th particle in a viscous carrier is additionally described by an analogous equation 
\begin{equation}
\frac{d \vec{v}^*_i}{d t^*} = \vec{f}^*_i - \gamma^{*T} \vec{v}^*_i + \vec{\eta}^{*T}_i.\label{eq:lang_tran}
\end{equation}

\noindent \hly{All simulation are performed using reduced quantities, 
denoted here with the asterisk.
They are formally introduced through the usage of three parameters:
the thermal energy $k_B T$,
the diameter of a single particle $d$,
and the mass of a single particle $\mathfrak{M}$.}
Specifically, 
$\vec{v}^*_i = \vec{v}_i \sqrt{\mathfrak{M} / k_B T}$ and
$\vec{\omega}^*_i = \vec{\omega}_i \sqrt{\mathfrak{M} d^2 / k_B T}$ 
are the reduced linear and angular velocities, correspondingly.
$J^* = J / \mathfrak{M} d^2$ is the reduced moment of inertia,
$\gamma^{*T} = \gamma^{T} \sqrt{d^2 / \mathfrak{M}k_B T}$ and 
$\gamma^{*R} = \gamma^{R} \sqrt{1 / d^2 \mathfrak{M} k_B T}$ 
are the reduced translational and rotational friction coefficients, 
$\vec{\eta}^{*R}_i$ and $\vec{\eta}^{*T}_i$ are the random force and torque,
that have zero mean values and satisfy the 
standard fluctuation-dissipation relationship~\cite{coffey2004langevin}
\begin{equation}
\langle \eta^{*T(R)}_{i \alpha}(t^*_1) \eta^{*T(R)}_{j \beta}(t^*_2) \rangle = 2 \gamma^{*T(R)}\delta_{\alpha \beta} \delta_{ij}\delta^*(t^*_1 - t^*_2), 	
\end{equation}
$\alpha$ and $\beta$ denote Cartesian vector components,
$\delta^*(t^*)$ is the Dirac delta function,
$\delta_{ij}$ is the Kronecker delta,
the reduced time is $t^* = t \sqrt{k_B T / \mathfrak{M} d^2}$.
$\vec{\tau}^*_i = \mu_0 \left[\vec{m}_i \times \left(\vec{H}_0 + \vec{H}_{dd}(i)\right)\right]/k_B T$ 
is the reduced magnetic torque acting on a given particle,
$\vec{H}_{dd}(i) = - (1/\mu_0)\sum_{j \neq i} \partial U_{dd}(i,j)/ \partial \vec{m}_i$ 
is the sum of all dipolar fields in the particle center,
$\vec{f}^*_i = - (d/k_B T)\sum_{j \neq i} \partial \left( U_{dd}(i,j) + U_{WCA}(i,j)\right)/ \partial \vec{r}_i$
is the total reduced force on the particle,
$U_{WCA}(i,j)$ is the Weeks-Chandler-Andersen (WCA) pair potential 
that models the steric repulsion between particles~\cite{weeks1971role}:
\begin{gather}  
		U_{WCA}(i,j) =
		\begin{cases}
		U_{LJ}(r_{ij}) - U_{LJ}(r_{cut}), & r_{ij} < r_{cut}  \\
		0, & r_{ij} \geq r_{cut},
		\end{cases}, \label{uwca} \\
		U_{LJ}(r) = 4 \varepsilon\left[\left(\frac{d}{r}\right)^{12} - \left(\frac{d}{r}\right)^{6}\right],
\end{gather}
where $U_{LJ}$ is the Lennard-Jones potential, $r_{cut} = 2^{1/6}d$.


The forces and torques due to long-range dipole-dipole interactions are computed using the dipolar P$^3$M algorithm with ``metallic'' boundary conditions~\cite{cerda2008p}.
All the results are reported for 
$J^* = \gamma^*_R = \gamma^*_T = \varepsilon^* = 1$,
simulation time step is $\Delta t^* = 0.01$.
Typically, the first $2 \times 10^5$ time steps are used for 
the system equilibration,
and the subsequent production run lasts for at least another 
$8 \times 10^5$ time steps.

\bibliography{references}

\end{document}